\title{Operational significance of the deviation equation in relativistic geodesy}

\author{Dirk Puetzfeld\footnote{Email: dirk.puetzfeld@zarm.uni-bremen.de, URL: http://puetzfeld.org} \\
        ZARM\\
        University of Bremen\\
        Am Fallturm, 28359 Bremen, Germany \and
        Yuri N.\ Obukhov\footnote{Email: obukhov@ibrae.ac.ru}\\
        Theoretical Physics Laboratory\\
        Nuclear Safety Institute, Russian Academy of Sciences\\
        B.Tulskaya 52, 115191 Moscow, Russia}

\date{}

\documentclass[11pt,english,twocolumn]{article}
\usepackage[]{graphicx}
\usepackage{amsmath,amsfonts,amssymb,natbib,makeidx,xcolor}
\bibpunct{(}{)}{,}{a}{}{;}
\makeindex
\begin{document}
\renewcommand{\abstractname}{Definitions}
\maketitle

\begin{abstract}
{\bf{Deviation equation}\index{equation!deviation}.}  Second order differential equation for the 4-vector which measures the distance between reference points on neighboring world lines in spacetime manifolds. 

\noindent {\bf{Relativistic geodesy}\index{relativistic!geodesy}.}  Science representing the Earth (or any planet), including the measurement of its gravitational field, in a four-dimensional curved spacetime using differential-geometric methods in the framework of Einstein's theory of gravitation (General Relativity).
\end{abstract}

\section*{Introduction}\label{sec_gr}

How does one measure the gravitational field in Einstein's theory? What is the foundation of relativistic gradiometry? The deviation equation gives answers to these fundamental questions. 

In Einstein's theory of gravitation, i.e.\ General Relativity\index{General Relativity}, the gravitational field manifests itself in the form of the Riemannian curvature tensor\index{tensor!curvature} $R_{abcd}$ \citep{Synge:1960}. This 4th-rank tensor can be defined as a measure of the noncommutativity of the parallel transport process of the underlying spacetime manifold \citep*{Synge:Schild:1978}. In terms of the covariant derivative\index{derivative!covariant} $\nabla_a$, and for an arbitrary tensor $T^{c_1 \dots c_k}{}_{d_1 \dots d_l}$, it is introduced via
\begin{eqnarray}
\left(\nabla_a\nabla_b - \nabla_b\nabla_a\right)T^{c_1 \dots c_k}{}_{d_1 \dots d_l} \nonumber \\
= \sum^{k}_{i=1} R_{abe}{}^{c_i} T^{c_1 \dots e \dots c_k}{}_{d_1 \dots d_l} \nonumber \\
- \sum^{l}_{j=1} R_{abd_j}{}^{e} T^{c_1 \dots c_k}{}_{d_1 \dots e \dots d_l}. \label{curvature_def}
\end{eqnarray}
General Relativity is formulated on a four-dimensional (pseudo) Riemannian spacetime\index{spacetime!Riemannian}\footnote{In our conventions the signature of the spacetime metric is assumed to be $(+1,-1,-1,-1)$}, and therefore the curvature tensor in Einstein's theory has twenty (20) independent components for the most general field configurations produced by nontrivial matter sources, whereas in vacuum the number of independent components reduces to ten (10). As compared to Newton's theory, the  gravitational field thus has more degrees of freedom in the relativistic framework. 

A central question in General Relativity, and consequently in relativistic geodesy, is how these components of the gravitational field\index{gravitational! field} can be determined in an operational way. Historically, \citet{Pirani:1956} was the first to point out that one could determine the full Riemann tensor\index{tensor!Riemann curvature} with the help of a (sufficiently large) number of test bodies in the vicinity of observer's world line. Pirani's suggestion to measure the curvature was based on the equation which describes the dynamics of a vector connecting two adjacent geodesics in spacetime. In the literature this equation is known as Jacobi equation\index{equation!Jacobi}, or geodesic deviation equation\index{equation!geodesic deviation}; its early derivations in a Riemannian context can be found in \citet{LeviCivita:1926,Synge:1926,Synge:1927}. 

A modern derivation and extension of the deviation equation, based on \citep*{Puetzfeld:Obukhov:2016:1}, is presented in the next section. In particular, it is explicitly shown, how a suitably prepared set of test bodies can be used to determine all components of the curvature of spacetime (and thereby to measure the gravitational field) with the help of an exact solution for the components of the Riemann tensor in terms of the mutual accelerations between the constituents of a cloud of test bodies and the observer.\index{observer} This can be viewed as an explicit realization of Szekeres' ``gravitational compass\index{gravitational!compass}'' \citep{Szekeres:1965}, or Synge's ``curvature detector'' \citep{Synge:1960}. In geodetic terms, such a solution represents a realization of a relativistic gradiometer\index{gradiometer!relativistic} or tensor gradiometer\index{gradiometer!tensor}, which has a direct operational relevance and forms the basis of relativistic gradiometry. 

\section*{Deviation equation}\label{sec_deviation_equation}

\begin{figure}
\begin{center}
\includegraphics[width=7cm]{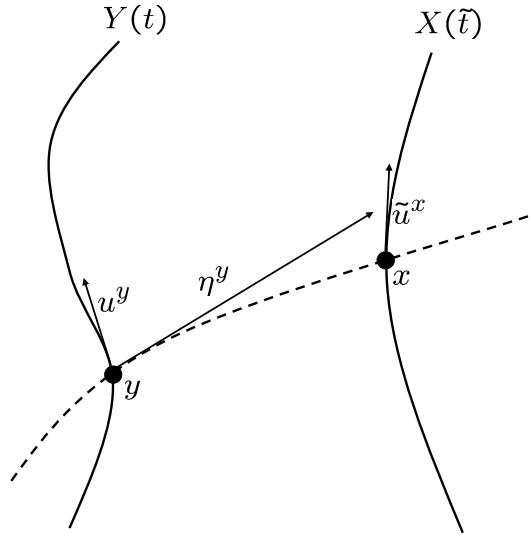}
\end{center}
\caption{\label{fig_setup} Sketch of the two arbitrarily parametrized world lines $Y(t)$ and $X(\tilde{t})$, and the geodesic\index{geodesic} connecting two points on these world line. The deviation vector along the reference world line $Y$ is denoted by $\eta^y$.}
\end{figure}

Let us consider two curves $Y(t)$ and $X(\tilde{t})$ in an arbitrary spacetime manifold\index{spacetime!manifold}, cf.\ fig.\ \ref{fig_setup}. They are not necessarily parameterized by the proper time and we allow for general parameters $t$ and $\tilde{t}$ along the curves. Any two points $x\in X$ and $y\in Y$ on the two curves are connected by the geodesic, which is unique provided the curves are sufficiently close. 

Along the connecting geodesic, the world function\index{world function} $\sigma(x,y)$ \citep{Synge:1960} is defined, which measures the finite distance between the spacetime points $x$ and $y$. By definition $\sigma(x,y)$ is a scalar function, which equals half the square geodesic distance between the points $x$ and $y$. Due to its dependence on two points, $\sigma(x,y)$ is also called biscalar, or two-point function.

The covariant derivative\index{derivative!covariant} of the world function\index{world function} $\sigma^y:=\nabla^y \sigma$ at $y$ is the conceptually closest object to the {\it connecting vector\index{connecting vector}} between the two points $y$ and $x$. To keep the formulas compact, it is convenient to suppress the tensor indices, and merely to display (as sub/superscripts) the spacetime point to which the suppressed index belongs. The higher order covariant derivatives of the world function are then denoted by $\sigma^y{}_{x_1\dots y_2\dots} := \nabla_{x_1}\dots\nabla_{y_2}\dots (\sigma^y)$. 

By construction, $\sigma^y$ is a tangent vector at the point $y$ with its length being the geodesic length between $y$ and $x$, and in flat spacetime it coincides with the connecting vector. With an account of these properties, one can infer a propagation equation for this ``generalized'' connecting vector along the reference curve\index{reference!curve}, cf.\ fig.~\ref{fig_setup}. Choosing $Y(t)$ as the reference curve, we define the generalized connecting vector to be:
\begin{eqnarray}
\eta^y := - \sigma^y. \label{gen_dev_definition} 
\end{eqnarray}
For its covariant total derivative, we have
\begin{eqnarray}
\frac{D}{dt} \eta^{y_1} &=& - \frac{D}{dt} \sigma^{y_1}\left(Y(t),X(\tilde{t})\right) \nonumber \\
&=& - \sigma^{y_1}{}_{y_2} \frac{\partial Y^{y_2}}{\partial t} - \sigma^{y_1}{}_{x_2} \frac{\partial X^{x_2}}{\partial \tilde{t}} \frac{d\tilde{t}}{dt} \nonumber \\
&=& - \sigma^{y_1}{}_{y_2} u^{y_2} - \sigma^{y_1}{}_{x_2} \tilde{u}^{x_2} \frac{d\tilde{t}}{dt}, \label{eta_1st_deriv}
\end{eqnarray}
where in the last line we defined the velocities $u^{y}:={\partial Y^{y}}/{\partial t}$ and $\tilde{u}^{x}:={\partial X^{x}}/{\partial \tilde{t}}$ along the two curves $Y$ and $X$, see fig.~\ref{fig_setup}. The second derivative of (\ref{eta_1st_deriv}) yields
\begin{eqnarray}
\frac{D^2}{dt^2} \eta^{y_1} &&= - \sigma^{y_1}{}_{y_2 y_3} u^{y_2} u^{y_3} - 2 \sigma^{y_1}{}_{y_2 x_3} u^{y_2} \tilde{u}^{x_3} \frac{d\tilde{t}}{dt} \nonumber \\
&& - \sigma^{y_1}{}_{y_2} a^{y_2} - \sigma^{y_1}{}_{x_2 x_3} \tilde{u}^{x_2} \tilde{u}^{x_3} \left(\frac{d\tilde{t}}{dt} \right)^2 \nonumber \\
&& - \sigma^{y_1}{}_{x_2} \tilde{a}^{x_2} \left(\frac{d\tilde{t}}{dt} \right)^2 - \sigma^{y_1}{}_{x_2} \tilde{u}^{x_2}  \frac{d^2\tilde{t}}{dt^2}, \label{eta_2nd_deriv}
\end{eqnarray}
here we introduced the accelerations for both curves, $a^y:={D u^y}/dt$, and $\tilde{a}^x:={D \tilde{u}^x}/d\tilde{t}$. 

In fact, equation (\ref{eta_2nd_deriv}) is already the generalized deviation equation, however, in order to give it an operational meaning all the quantities should be defined along the reference word line $Y$ -- along which the observer\index{observer} moves and performs his measurements. This is achieved by performing covariant expansions of all quantities around the reference world line\footnote{The technical details of the covariant Taylor expansion technique can be found in \citep*{Puetzfeld:Obukhov:2016:1}.}, which results in the expanded version of (\ref{eta_2nd_deriv}) in powers of the world function. Up to the second order we have: 
\begin{eqnarray}
 \frac{D^2}{dt^2} \eta^{y_1} &=& \tilde{a}^{y_1} \left(\frac{d\tilde{t}}{dt} \right)^2 - a^{y_1}\nonumber \\ &+& \frac{d t}{d \tilde{t}} \frac{d^2\tilde{t}}{dt^2} u^{y_1}  + \frac{D \eta^{y_1}}{dt} \frac{d t}{d \tilde{t}} \frac{d^2\tilde{t}}{dt^2} \nonumber \\
&-&\eta^{y_4} R^{y_1}{}_{y_2 y_3 y_4} \bigg(  u^{y_2} u^{y_3}  + 2 u^{y_3} \frac{D \eta^{y_2}}{dt} \bigg)\nonumber \\ 
&+& \mathcal{O}(\sigma^2). \label{eta_2nd_deriv_compact_2}
\end{eqnarray}
This {\it deviation equation\index{equation!deviation}} describes the change of the connecting vector $\eta^y$ between two general world lines. It is valid for completely general parametrizations of the curves $Y$ and $X$. 

Equation (\ref{eta_2nd_deriv_compact_2}) can be used to operationally model the relative motion of two objects -- in the context of relativistic geodesy one may think of two satellites -- which are subject to gravitational as well as other physical forces. The external, non-gravitational, forces are represented by the accelerations $\tilde{a}^{y}$ and $a^{y}$ in (\ref{eta_2nd_deriv_compact_2}). The gravitational forces are encoded in the curvature tensor $R^{y_1}{}_{y_2 y_3 y_4}$. This fact can be utilized to develop a measurement procedure for the gravitational field by means of the deviation equation (\ref{eta_2nd_deriv_compact_2}).

\section*{Measuring the gravitational field}\label{sec_deviation_equation2}

\begin{figure}
\begin{center}
    \includegraphics[width=8.5cm,angle=-90]{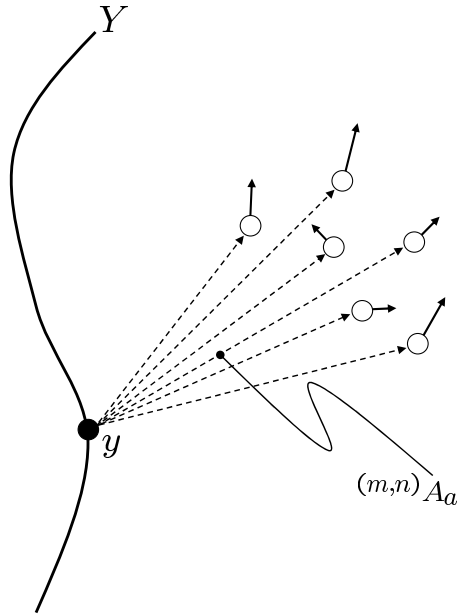}
\end{center}
   \caption{\label{fig_compass_sketch} Sketch of the operational procedure to measure the curvature of spacetime. An observer\index{observer} moving along a world line $Y$ monitors the accelerations ${}^{(m,n)}A_a$ to a set of suitably prepared test bodies (hollow circles). The number of test bodies required for the determination of all curvature components depends on the type of the underlying spacetime.} 
\end{figure}

The operational procedure, see fig.\ \ref{fig_compass_sketch}, is to monitor the accelerations of a set of test bodies in the vicinity of the observer\index{observer} who moves along the reference world line\index{reference!world line} $Y$. A mechanical analogue would be to measure the forces between the test bodies and the observer via interconnecting springs. 

Being interested in the gravitational field, we assume that the external accelerations in (\ref{eta_2nd_deriv_compact_2}) vanish. Furthermore, we assume that we are dealing with structureless test bodies, in other words the world lines $Y$ and $X$ become geodesics. With the additional choice of a synchronous parameterization of the world lines, see \citep*{Puetzfeld:Obukhov:2016:1} for details, the general deviation equation (\ref{eta_2nd_deriv_compact_2}) then turns into the {\it geodesic deviation, or Jacobi, equation}:
\begin{eqnarray}
\frac{D^2}{ds^2}\eta^a = R^a{}_{bcd} u^b \eta^c u^d. \label{compass_start}
\end{eqnarray}
Here $s$ denotes the proper time along the reference curve. In order to exploit this equation operationally, the covariant derivative of the deviation vector $\eta^a$ needs to be rewritten in terms of the standard (non-covariant) derivative. This can be achieved by employing normal coordinates\index{normal coordinates} along the world line of the observer \citep*{Puetzfeld:Obukhov:2016:1}, in which the Levi-Civita connection\index{Levi-Civita!connection} $\Gamma_{ab}{}^c$ and its first derivative take the form
\begin{eqnarray}
\Gamma_{ab}{}^c |_Y= 0, \quad \quad \partial_a \Gamma_{bc}{}^d|_Y = \frac{2}{3} R_{a(bc)}{}^d,
\end{eqnarray}
along $Y$. In the normal coordinates, the deviation equation (\ref{compass_start}) is recast into
\begin{eqnarray}
\frac{d^2}{ds^2}\eta_a &\stackrel{|_Y}{=}& \frac{4}{3} R_{abcd} u^b \eta^c u^d. \label{compass_lower}
\end{eqnarray} 
This equation has the formal structure
\begin{eqnarray}
{\rm acceleration \quad} &=& {\rm \quad gravitational \,\,  field \quad} \times \nonumber \\
&&{\rm \quad position \quad } \times {\rm \quad velocity}^2, \nonumber
\end{eqnarray}
and it allows to express the curvature in terms of measured and/or prescribed quantities. Equation (\ref{compass_lower}) forms the basis for setting up a gravitational compass or relativistic gradiometer by means of free falling test bodies.

\section*{Gravitational compass (Relativistic gradiometer)}\label{sec_compass_solution}

Recalling the idea of \citet{Pirani:1956}, we now set up a cloud of test bodies in the vicinity of the observer\index{observer}. The goal is to find a configuration of test bodies, which allows for a complete determination of the gravitational field.   

For $(n)$ bodies at locations ${}^{(n)}\eta^a$ relative to the reference body, moving with relative $(m)$ velocities ${}^{(m)}u^a$ we end up with the system
\begin{eqnarray}
{}^{(m,n)}A_a &\stackrel{|_Y}{=}& \frac{4}{3} R_{abcd} {}^{(m)}u^b \, {}^{(n)}\eta^c \, {}^{(m)}u^d. \label{compass_lower_setup}
\end{eqnarray}
Here we denote by ${}^{(m,n)}A_a$ the measured accelerations relative to the reference point $Y$ of the individual test bodies. Physically, these $A$'s correspond to the springs in the mechanical compass picture of \citet{Szekeres:1965}.

\begin{figure}
\begin{center}
\includegraphics[width=5cm,angle=-90]{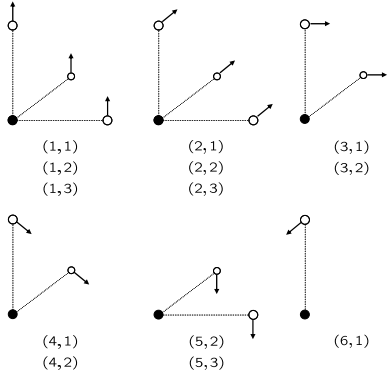} 
\end{center}
\caption{\label{fig_standard_compass_general} Symbolical sketch of the explicit general compass solution in (\ref{exR1010})-(\ref{ex3212}). In total 13 suitably prepared test bodies (hollow circles) are needed to determine all 20 curvature components. The observer\index{observer} is denoted by the black circle. With the standard deviation equation ${}^{(1 \dots 6)}u^{a}$, as well as ${}^{(1 \dots 3)}\eta^{a}$ are needed in the solution.}
\end{figure}

In a general spacetime, all 20 independent components of the curvature tensor are determined in terms of the accelerations ${}^{(m,n)}A_a$ and velocities ${}^{(m)}u^a$, if we use the setup sketched in fig.\ \ref{fig_standard_compass_general}. This can be achieved with the help of suitably prepared test bodies at locations 
\begin{eqnarray}
&&{}^{(1)}\eta^{a}=\left(\begin{array}{c} 0 \\ 1\\ 0\\ 0\\ \end{array} \right),
{}^{(2)}\eta^{a}=\left(\begin{array}{c} 0 \\ 0\\ 1\\ 0\\ \end{array} \right), \nonumber \\
&&{}^{(3)}\eta^{a}=\left(\begin{array}{c} 0 \\ 0\\ 0\\ 1\\ \end{array} \right), \label{position_setup}
\end{eqnarray}
with velocities 
\begin{eqnarray}
&&{}^{(1)}u^{a}=\left(\begin{array}{c} c_{10} \\ 0 \\ 0\\ 0\\ \end{array} \right),
{}^{(2)}u^{a}=\left(\begin{array}{c}  c_{20} \\ c_{21} \\ 0 \\ 0\\ \end{array} \right), \nonumber \\
&&{}^{(3)}u^{a}=\left(\begin{array}{c}  c_{30} \\ 0 \\ c_{32} \\ 0 \\ \end{array} \right),
{}^{(4)}u^{a}=\left(\begin{array}{c}  c_{40} \\ 0 \\ 0 \\ c_{43}\\ \end{array} \right), \nonumber \\
&&{}^{(5)}u^{a}=\left(\begin{array}{c}  c_{50} \\ c_{51} \\ c_{52} \\ 0\\ \end{array} \right),
{}^{(6)}u^{a}=\left(\begin{array}{c}  c_{60} \\ 0 \\ c_{62} \\ c_{63} \\ \end{array} \right). 
\label{velocity_setup}
\end{eqnarray}
From the algebraic system (\ref{compass_lower_setup}), we find the components of the Riemann curvature\index{curvature!Riemann}\index{tensor!Riemann curvature} tensor in terms of the velocity components $c_{mi}$ (where $m = 1,\dots, 6$ and $i= 0,1,2,3$) and the accelerations 
\begin{eqnarray}
{}^{(1,1)}\!\!A_1, {}^{(1,1)}\!\!A_2, {}^{(1,1)}\!\!A_3, {}^{(1,2)}\!\!A_2, {}^{(1,2)}\!\!A_3, {}^{(1,3)}\!\!A_3, \nonumber\\
{}^{(2,1)}\!\!A_2, {}^{(2,1)}\!\!A_3, {}^{(2,2)}\!\!A_2, {}^{(2,2)}\!\!A_3, {}^{(2,3)}\!\!A_3, \nonumber\\
{}^{(4,1)}\!\!A_0, {}^{(4,1)}\!\!A_2, {}^{(4,1)}\!\!A_3, {}^{(4,2)}\!\!A_0, {}^{(4,2)}\!\!A_2, \nonumber\\
{}^{(3,1)}\!\!A_0, {}^{(3,2)}\!\!A_3, {}^{(5,2)}\!\!A_3, {}^{(5,3)}\!\!A_3, {}^{(6,1)}\!\!A_1. 
\end{eqnarray}
This implies that 13 test bodies\index{test bodies} are needed to measure the gravitational field completely. The explicit solution is given in appendix \ref{app_explicit_sols_gen}, and is represented in graphical form in fig.\ \ref{fig_standard_compass_general}. 

\section*{Vacuum solution}\label{sec_vacuum_solution}

In vacuum the number of independent components of the curvature tensor is reduced to the 10 components of the Weyl tensor\index{curvature!Weyl}\index{tensor!Weyl curvature} $C_{abcd}$. Replacing $R_{abcd}$ in the compass solution and taking into account the symmetries of Weyl tensor (in particular, the double-self-duality property $C_{abcd} = -\frac{1}{4}\epsilon_{abef}\epsilon_{cdgh}C^{efgh}$, where $\epsilon_{abcd}$ is the totally antisymmetric Levi-Civita tensor\index{tensor!Levi-Civita}\index{Levi-Civita!tensor} with $\epsilon_{0123}=1$), we may use a reduced compass setup to completely determine the 10 vacuum components of the gravitational field in terms of the accelerations
\begin{eqnarray}
{}^{(1,1)}\!\!A_1, {}^{(1,1)}\!\!A_2, {}^{(1,1)}\!\!A_3, {}^{(1,2)}\!\!A_2, {}^{(1,2)}\!\!A_3, \nonumber\\
{}^{(2,1)}\!\!A_2, {}^{(2,1)}\!\!A_3, {}^{(2,2)}\!\!A_3, {}^{(3,1)}\!\!A_0, {}^{(4,1)}\!\!A_2.
\end{eqnarray}
This implies that one needs 6 test bodies\index{test bodies} to measure the gravitational field in vacuum. The explicit solution is given in appendix \ref{app_explicit_sols_vac}. 

This completes the construction of a gravitational compass\index{gravitational!compass} \citep{Szekeres:1965}, or relativistic gradiometer, on the basis of the geodesic deviation equation (\ref{compass_lower}). 

We only note in passing, that current research indicates a possible reduction of the number of required test bodies if one makes use of the generalized deviation equation\index{equation!generalized deviation} (\ref{eta_2nd_deriv}). A detailed discussion and a comparison to other explicit compass solutions in the literature \citep*{Ciufolini:Demianski:1986} can be found in \citep*{Puetzfeld:Obukhov:2016:1}.  

\begin{figure}
\begin{center}
\includegraphics[width=5cm,angle=-90]{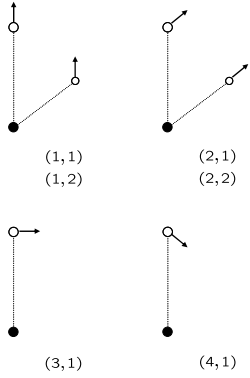}
\end{center}
\caption{\label{fig_standard_compass_vacuum} Sketch of the explicit compass solution in (\ref{exC1010})-(\ref{exC0312}) for the vacuum case. In total 6 suitably prepared test bodies\index{test bodies} (hollow circles) are needed to determine all 10 components of the Weyl tensor\index{curvature!Weyl}\index{tensor!Weyl}. The observer is denoted by the black circle. With the standard deviation equation, only ${}^{(1 \dots 4)}u^{a}$, as well as ${}^{(1 \dots 2)}\eta^{a}$ are needed in the solution.}
\end{figure}

\section*{Summary}\label{sec_summary}

Deviation equations form the theoretical basis for many experiments. They provide the foundation of relativistic gradiometry\index{relativistic!gradiometry}, and allow for the determination of the gravitational field (the curvature) by means of acceleration measurements in the vicinity of an observer. 

The minimal number of test bodies\index{test bodies} required to establish all components of gravitational field can be obtained with the help of deviation equations. In particular, one may use such equations to develop explicit detector setups for relativistic gradiometers.

In relativistic geodesy\index{relativistic!geodesy}, based on Einstein's theory of General Relativity, one needs at least 13 test bodies to determine all gravitational field components in a general spacetime. In a vacuum spacetimes\index{spacetime!vacuum} this number is reduced to 6. 

\section*{Acknowledgements}
This work was supported by the Deutsche Forschungsgemeinschaft (DFG) through the grant PU 461/1-1 (D.P.). The work of Y.N.O. was partially supported by PIER (``Partnership for Innovation, Education and Research'' between DESY and Universit\"at Hamburg) and by the Russian Foundation for Basic Research (Grant No. 16-02-00844-A). 

\appendix
\section{Explicit solutions}\label{app_explicit_sols}
\subsection{General spacetime}\label{app_explicit_sols_gen}
\begin{eqnarray}
01 : R_{1010} &=& \frac{3}{4} {}^{(1,1)}A_1 c^{-2}_{10}, \label{exR1010} \\
02 : R_{2010} &=& \frac{3}{4} {}^{(1,1)}A_2 c^{-2}_{10}, \label{exR2010} \\
03 : R_{3010} &=& \frac{3}{4} {}^{(1,1)}A_3 c^{-2}_{10}, \label{exR3010} \\
04 : R_{2020} &=& \frac{3}{4} {}^{(1,2)}A_2 c^{-2}_{10}, \label{exR2020} \\
05 : R_{3020} &=& \frac{3}{4} {}^{(1,2)}A_3 c^{-2}_{10}, \label{exR3020} \\
06 : R_{3030} &=& \frac{3}{4} {}^{(1,3)}A_3 c^{-2}_{10}, \label{exR3030} \\
07 : R_{2110} &=& \frac{3}{4} {}^{(2,1)}A_2 c^{-1}_{21} c^{-1}_{20} -  R_{2010} c^{-1}_{21} c_{20} ,  \nonumber \\ 
\label{exR2110} \\
08 : R_{3110} &=& \frac{3}{4} {}^{(2,1)}A_3 c^{-1}_{21} c^{-1}_{20} -  R_{3010} c^{-1}_{21} c_{20} , \nonumber \\
\label{exR3110}\\
09 : R_{0212} &=& \frac{3}{4} {}^{(3,1)}A_0 c^{-2}_{32} +  R_{2010} c^{-1}_{32} c_{30} , \label{exR0212}\\
10 : R_{1212} &=& \frac{3}{4} {}^{(2,2)}A_2 c^{-2}_{21} -  R_{2020} c^{2}_{20} c^{-2}_{21}  \nonumber \\
 &&-  2 R_{0212} c^{-1}_{21} c_{20}  , \label{exR1212}\\
11 : R_{3220} &=& \frac{3}{4} {}^{(3,2)}A_3 c^{-1}_{32} c^{-1}_{30} -  R_{3020} c^{-1}_{32} c_{30} ,  \nonumber \\ 
\label{exR3220}\\
12 : R_{0313} &=& \frac{3}{4} {}^{(4,1)}A_0 c^{-2}_{43} +  R_{3010} c^{-1}_{43} c_{40} , \label{exR0313} \\
13 : R_{1313} &=& \frac{3}{4} {}^{(2,3)}A_3 c^{-2}_{21} -  R_{3030} c^{2}_{20} c^{-2}_{21}   \nonumber \\
 && -  2 R_{0313} c^{-1}_{21} c_{20}  , \label{exR1313} \\
14 : R_{0323} &=& \frac{3}{4} {}^{(4,2)}A_0 c^{-2}_{43} +  R_{3020} c^{-1}_{43} c_{40} , \label{exR0323}\\
15 : R_{2323} &=& \frac{3}{4} {}^{(4,2)}A_2 c^{-2}_{43} -  R_{2020} c^{-2}_{43} c^{2}_{40}   \nonumber \\
 && +  2 R_{3220} c^{-1}_{43} c_{40}  , \label{exR2323} 
\end{eqnarray}
\begin{eqnarray}
16 : R_{3132} &=& \frac{3}{8} {}^{(5,3)}A_3 c^{-1}_{52} c^{-1}_{51}  -  \frac{1}{2} R_{3030} c^{-1}_{52} c^{-1}_{51} c^2_{50} \nonumber \\ 
&&  -  R_{0313} c^{-1}_{52} c_{50} -  R_{0323} c^{-1}_{51} c_{50}  \nonumber \\
&&  -  \frac{1}{2} R_{1313} c^{-1}_{52} c_{51} - \frac{1}{2} R_{2323} c_{52} c^{-1}_{51} , \nonumber \\
 \label{exR3132} \\
17 : R_{1213} &=& \frac{3}{8} {}^{(6,1)}A_1 c^{-1}_{63} c^{-1}_{62}  -  \frac{1}{2} R_{1010} c^{-1}_{63} c^{-1}_{62} c^2_{60} \nonumber \\ 
&&  +  R_{2110} c^{-1}_{63} c_{60} +  R_{3110} c^{-1}_{62} c_{60} \nonumber \\
&&   -  \frac{1}{2} R_{1212} c^{-1}_{63} c_{62} - \frac{1}{2} R_{1313} c_{63} c^{-1}_{62} , \nonumber \\
 \label{exR1213}\\
18 : R_{0231} &=& \frac{1}{4} {}^{(4,1)}A_2 c^{-1}_{40} c^{-1}_{43}  -  \frac{1}{4} {}^{(2,2)}A_3 c^{-1}_{20} c^{-1}_{21} \nonumber \\ 
&&  + \frac{1}{3} \big( R_{3020} c_{20} c^{-1}_{21} + R_{3121} c_{21} c^{-1}_{20} \nonumber \\
&&  -  R_{2010} c_{40} c^{-1}_{43} -  R_{2313} c_{43} c^{-1}_{40} \big),  \label{ex0231}\\ 
19 : R_{0312} &=& \frac{1}{4} {}^{(4,1)}A_2 c^{-1}_{40} c^{-1}_{42}  + \frac{1}{2} {}^{(2,2)}A_3 c^{-1}_{20} c^{-1}_{21} \nonumber \\ 
&&  - \frac{1}{3} \big( 2 R_{3020} c_{20} c^{-1}_{21} + 2 R_{3121} c_{21} c^{-1}_{20} \nonumber \\
&&  + R_{2010} c_{40} c^{-1}_{43} +  R_{2313} c_{43} c^{-1}_{40}  \big),  \label{ex0312}\\
20 : R_{3212} &=& \frac{3}{4} {}^{(4,1)}A_3  c^{-1}_{20}  c^{-1}_{21}  c_{50}  c^{-1}_{52}  \nonumber \\
&& -  \frac{3}{4} {}^{(5,2)}A_3  c^{-1}_{51}  c^{-1}_{52}  \nonumber \\ 
&& + R_{3121}  c^{-1}_{52} \left(c_{51} - c_{50}  c_{21} c^{-1}_{20}\right) \nonumber \\
&& + R_{3220} c_{50} c^{-1}_{51} \nonumber \\
&&  + R_{3020}  c_{50}  c^{-1}_{52} \left(c_{50} c^{-1}_{51} -c_{20}  c^{-1}_{21} \right) . \nonumber \\
 \label{ex3212}
\end{eqnarray}

\subsection{Vacuum spacetime}\label{app_explicit_sols_vac}

\begin{eqnarray}
01 : C_{1010} &=& \frac{3}{4} {}^{(1,1)}A_1 c^{-2}_{10}, \label{exC1010} \\
02 : C_{2010} &=& \frac{3}{4} {}^{(1,1)}A_2 c^{-2}_{10}, \label{exC2010} \\
03 : C_{3010} &=& \frac{3}{4} {}^{(1,1)}A_3 c^{-2}_{10}, \label{exC3010} 
\end{eqnarray}
\begin{eqnarray}
04 : C_{2020} &=& \frac{3}{4} {}^{(1,2)}A_2 c^{-2}_{10}, \label{exC2020} \\
05 : C_{3020} &=& \frac{3}{4} {}^{(1,2)}A_3 c^{-2}_{10}, \label{exC3020} \\
06 : C_{2110} &=& \frac{3}{4} {}^{(2,1)}A_2 c^{-1}_{21} c^{-1}_{20} -  C_{2010} c^{-1}_{21} c_{20} , \nonumber \\
\label{exC2110}\\
07 : C_{3110} &=& \frac{3}{4} {}^{(2,1)}A_3 c^{-1}_{21} c^{-1}_{20} -  C_{3010} c^{-1}_{21} c_{20} , \nonumber \\
\label{exC3110} \\
08 : C_{0212} &=& \frac{3}{4} {}^{(3,1)}A_0 c^{-2}_{32} +  C_{2010} c^{-1}_{32} c_{30} , \label{exC0212} \\ 
09 : C_{0231} &=& \frac{1}{4} {}^{(4,1)}A_2 c^{-1}_{40} c^{-1}_{43}  -  \frac{1}{4} {}^{(2,2)}A_3 c^{-1}_{20} c^{-1}_{21} \nonumber \\ 
&&  + \frac{1}{3} C_{3020} \big(  c_{20} c^{-1}_{21} + c_{21} c^{-1}_{20}\big) \nonumber \\
&&  - \frac{1}{3} C_{2010}\big( c_{40} c^{-1}_{43} + c_{43} c^{-1}_{40} \big),  \label{exC0231} \\ 
10 : C_{0312} &=& \frac{1}{4} {}^{(4,1)}A_2 c^{-1}_{40} c^{-1}_{42}  + \frac{1}{2} {}^{(2,2)}A_3 c^{-1}_{20} c^{-1}_{21} \nonumber \\ 
&&  - \frac{2}{3} C_{3020} \big( c_{20} c^{-1}_{21} + c_{21} c^{-1}_{20} \big) \nonumber \\
&&  + \frac{1}{3} C_{2010} \big( c_{40} c^{-1}_{43} + c_{43} c^{-1}_{40} \big). \label{exC0312}
\end{eqnarray}

\bibliographystyle{plainnat}
\bibliography{encyclogeo_puetzfeld_obukhov}

\begin{thebibliography}{9}
\providecommand{\natexlab}[1]{#1}
\providecommand{\url}[1]{\texttt{#1}}
\expandafter\ifx\csname urlstyle\endcsname\relax
  \providecommand{\doi}[1]{doi: #1}\else
  \providecommand{\doi}{doi: \begingroup \urlstyle{rm}\Url}\fi

\bibitem[{Ciufolini} and {Demianski}(1986)]{Ciufolini:Demianski:1986}
I.~{Ciufolini} and M.~{Demianski}.
\newblock {How to measure the curvature of space-time}.
\newblock \emph{Phys. Rev. D}, 34:\penalty0 1018, 1986.

\bibitem[{Levi-Civita}(1926)]{LeviCivita:1926}
T.~{Levi-Civita}.
\newblock {Sur l'{}\'ecart g\'eod\'esique}.
\newblock \emph{Math. Ann.}, 97:\penalty0 291, 1926.

\bibitem[{Pirani}(1956)]{Pirani:1956}
F.~A.~E. {Pirani}.
\newblock {On the physical significance of the Riemann tensor}.
\newblock \emph{Acta Phys. Pol.}, 15:\penalty0 389, 1956.

\bibitem[{Puetzfeld} and {Obukhov}(2016)]{Puetzfeld:Obukhov:2016:1}
D.~{Puetzfeld} and Yu.~N. {Obukhov}.
\newblock {Generalized deviation equation and determination of the curvature in
  General Relativity}.
\newblock \emph{Phys. Rev. D}, 93:\penalty0 044073, 2016.

\bibitem[{Synge}(1926)]{Synge:1926}
J.~L. {Synge}.
\newblock {The first and second variations of the length integral in Riemannian
  space}.
\newblock \emph{Proc. Lond. Math. Soc.}, 25:\penalty0 247, 1926.

\bibitem[{Synge}(1927)]{Synge:1927}
J.~L. {Synge}.
\newblock {On the geometry of dynamics}.
\newblock \emph{Phil. Trans. R. Soc. Lond. A}, 226:\penalty0 31, 1927.

\bibitem[{Synge}(1960)]{Synge:1960}
J.~L. {Synge}.
\newblock \emph{{Relativity: The general theory}}.
\newblock North-Holland, Amsterdam, 1960.

\bibitem[{Synge} and {Schild}(1978)]{Synge:Schild:1978}
J.~L. {Synge} and A.~{Schild}.
\newblock \emph{{Tensor calculus}}.
\newblock Dover, New York, 1978.

\bibitem[{Szekeres}(1965)]{Szekeres:1965}
P.~{Szekeres}.
\newblock {The gravitational compass}.
\newblock \emph{J. Math. Phys.}, 6:\penalty0 1387, 1965.

\end{thebibliography}

\end{document}